\documentclass[epj,final]{svjour}

\usepackage{latexsym}
\usepackage{url}
\usepackage{amsfonts}
\usepackage{amsmath, amssymb}
\RequirePackage{graphicx}

\def\d{{\nabla}}

\begin{document}
\title{Cosmic voids and induced hyperbolicity}
\author{M.~Samsonyan\inst{1}, A.A.~Kocharyan\inst{2}, A.~Stepanian\inst{1}, V.G.~Gurzadyan\inst{1,3}
}                     
%
%
\institute{Center for Cosmology and Astrophysics, Alikhanian National Laboratory and Yerevan State University, Yerevan, Armenia  \and 
School of Physics and Astronomy, Monash University, Clayton, Australia \and
SIA, Sapienza Universita di Roma, Rome, Italy}
\date{Received: date / Revised version: date}
%

\abstract{Cosmic voids - the low density regions in the Universe - as characteristic features of the large scale matter distribution, are known for their hyperbolic properties. The latter implies the deviation of photon beams due to their underdensity, thus mimicing the negative curvature. We now show that the hyperbolicity can be induced not only by negative curvature or underdensity but also depends on the anisotropy of the photon beams. } 
%
\maketitle

\section{Introduction}

The low density regions in the large scale Universe - voids - are among actively studied phenomena, see \cite{P1,P2,Ham,Din} and references therein. Cosmic voids are acting as probes for modified gravity theories, evolution of cosmological density perturbations, etc. Various observational surveys aim to reveal the characteristics of the voids (e.g.\cite{Hoy,Ce,Nad}), the distributions of their spatial scales, underdensity parameter, as their knowledge is of particular importance for the reconstruction of the spectrum of the density perturbations and the formation of the large scale Universe. 

Cosmic Microwave Background (CMB) provided another window to trace the presence of the voids \cite{Hig1,Hig2,Rag,Vi}, along with the traditional galaxy surveys. For example, the Cold Spot, a remarkable non-Gaussian feature known in the CMB sky was shown to reveal properties of a void \cite{spot}, as supported also with galactic survey \cite{Sz}. 

At the study of the Cold Spot the hyperbolicity property of voids was used, namely, the deviation of the photon trajectories, i.e. of null geodesics due to the underdensity of the void. The deviation of geodesic flows is known to be a property of negatively curved spaces as studied in theory of dynamical systems \cite{An,Arn}. Regarding the voids, it was shown that the low-density spatial regions can
induce hyperbolicity even in conditions of globally flat or positively curved Universe \cite{GK1,GK2}. The voids as divergent lenses were considered also in \cite{Das}.
        
Below we show that the hyperbolicity of geodesic flows can be caused not only by the underdensity parameter of a void but also will depend on the anisotropy of the photon beams. 

The property of hyperbolicity can be defined by means of the equation of deviation of close geodesics defined in a d-dim Riemannian manifold $M$, known as Jacobi equation  \cite{An,GK1}
\begin{equation}
 \d^2_un+\Re_u(n)=0,
\end{equation}
where the deviation vector 
$$n=\ell\hat{n}$$ 
is orthogonal to the velocity vector $u$ and $g(n,n)=\ell^2$, $g(\hat{n},\hat{n})=1$, $g(\hat{n},u)=0$. 

Jacobi equation can be written in the form 
\begin{equation}
  \ddot{\ell}+[K(u,\hat{n})-g(\d_u\hat{n},\d_u\hat{n})]\ell=0,\label{lOriginal}
\end{equation}
with the sectional curvature 
\begin{equation}
 K(u,\hat{n})=g(\Re_u(\hat{n}),\hat{n})=g(\Re(\hat{n},u)u,\hat{n}).
\end{equation}
For compact manifold $M$ and when 
\begin{equation}
K(u,n)<K_0<0
\end{equation}
for all orthonormal vectors $u$ and $n$ at any point of $M$, the geodesic flow is an Anosov system \cite{An}, so that the close geodesics deviate exponentially at any point of $M$ and at any two-dimensional directions defined by the two vectors $u$ and $n$. 
   		
Below, first, we will consider how a distortion parameter can be defined to enable the sought angular dependence on the geodesics deviation. Then, we will illustrate quantitatively the role of the angular dependence vs other parameters, i.e. the sign of the mean density of the medium containing the underdense void. We also mention the possible links to the observed effects where the considered hyperbolicity can have contribution.

\section{Distortion parameter}

For every (d+1)-dimensional Lorentzian manifold, the averaged geodesic deviation equation is \cite{GK1}
\begin{equation}\label{gdev}
\frac{d^2\ell}{d \eta^2}+ \frac{\mathfrak{r}}{d-1}\,\ell =0\,,
\end{equation}
where $\mathfrak{r}$ is the d-dimensional spatial Ricci scalar (the details of justification for these averages can be found in \cite{GK1}). On the other hand, for FLRW  metric with small perturbation $\phi$, the line element is written as ($c=1$)
\begin{equation}\label{FLRW}
ds^2 = -(1+2\phi) dt^2 + (1- 2\phi) a^2(t) d\gamma^2\,, 
\end{equation}
where depending on the sign of sectional curvature  $k$ of spatial geometry, $d\gamma^2$ represents the spherical ($k=1$), Euclidean ($k=0$) or hyperbolic ($k=-1$) geometries. Meantime, the perturbation field $\phi$, defined over the above metric, satisfies the following conditions \cite{Holz}
\begin{equation}\label{Pert}
|\phi|\ll1, \quad \left(\frac{\partial \phi}{\partial t}\right)^2\ll a^{-2} ||\nabla\phi||^2\, .
\end{equation}

Now if we take Eq.(\ref{FLRW}) as the weak-field limit, the special Ricci scalar will be ($d=3$)
\begin{equation}\label{RicciL}
\frac{\mathfrak{r}}{2}
= k + 2 H_0^2\Omega_{m}\ \Theta\ \tilde{\delta},
\end{equation}
where 
$$
\tilde{\delta}=\frac{\delta\rho}{a\langle\rho\rangle}
=\frac{\rho-\langle\rho\rangle}{a\langle\rho\rangle}
$$ 
and 
\begin{equation}\label{theta}
\Theta= \frac{3}{4}\left(1+ \frac{(\nabla^2\phi,\mathbf{u}\otimes\mathbf{u})}{\Delta\phi}\right)\,.
\end{equation}
$\Theta$ reflects the anisotropy of photon beams, $\Theta=1$ corresponding to a spherical distribution i.e. $\langle\mathbf{u}\otimes\mathbf{u}\rangle=\frac{1}{3}\mathbf{\gamma}$.

\noindent 
The importance of Eq.(\ref{RicciL}) lies on the fact that, it enables us to study the stability conditions for large cosmic structures i.e. voids and walls \cite{GK2}. In this sense, Eq.(\ref{gdev}) for our case will be written as
\begin{equation}
\frac{d^2\ell}{d \eta^2}+ \frac{\mathfrak{r}}{2}\ell =0\,,
\end{equation}

If $k=0$, then we have
\begin{equation}
\frac{d^2\ell}{d \eta^2}+ 2 H_0^2\Omega_{m}\,\Theta\,\tilde{\delta}\,\ell =0\,,
\end{equation}
or equivalently
\begin{equation}
\frac{d^2\ell}{d \tau^2}+ \Theta\,\tilde{\delta}\,\ell =0\,,
\end{equation}
where 
\begin{equation}
d\tau=H_0\sqrt{2\Omega_{m}}d\eta
=-\frac{\sqrt{2\Omega_{m}}\mathbf{d}z}{\sqrt{\Omega_\Lambda+\left[\Omega_k+\Omega_m(1+z)\right](1+z)^2}}
=-\frac{\sqrt{2}\mathbf{d}z}{\sqrt{\Omega_m^{-1}-1+(1+z)^3}}.
\end{equation}
As in \cite{GK2}, by adopting periodicity in the line-of-sight distribution of voids, i.e. 
\begin{equation}\label{period1}
\tilde{\delta}(\tau+\tau_k+\tau_\omega)=\tilde{\delta}(\tau) =
    \begin{cases}
      -\kappa^2 & 0< \tau < \tau_\kappa\\
      \ \omega^2 & \tau_\kappa < \tau < \tau_\kappa+\tau_\omega\, ,
    \end{cases}    
\end{equation}
and
$$
\Theta=1+\nu(\tau)\,,
$$
where $\nu$ is a stationary process with $\langle\nu(\tau)\rangle=0$ and auto-correlation function $\Gamma(\tau)=\langle\nu(\tilde\tau)\nu(\tilde\tau+\tau)\rangle=\sigma^2\delta(\tau)$. 

This leads to the following matrix equation:

\begin{equation}\label{average2}
    \frac{d}{d\tau}\begin{pmatrix}
\langle \ell^2\rangle\\
\langle\dot{\ell}^2\rangle\\
\langle\ell\dot\ell\rangle
\end{pmatrix}
=
\begin{pmatrix}
0 & 0 & 2\\
\sigma^2\tilde\delta^2 & 0 & -2\tilde\delta\\
-\tilde\delta & 1 & 0
\end{pmatrix}
\begin{pmatrix}
\langle\ell^2\rangle\\
\langle\dot{\ell}^2\rangle\\
\langle\ell\dot\ell\rangle
\end{pmatrix}\,.
\end{equation}
The matrix of transformation $f$ after period $\tau_\kappa+\tau_\omega$ for the system (\ref{average2}) is given by
\begin{equation}
f(\alpha)=e^{B\tau_\omega}e^{A\tau_\kappa}\left(\text{\it I}+ \alpha
\left[\kappa^4\int_0^{\tau_\kappa}e^{-As}Je^{As}ds
       +\omega^4 e^{-A\tau_\kappa}\left(\int_0^{\tau_\omega}e^{-Bs}Je^{Bs}ds\right)e^{A\tau_\kappa}\right]\right) + o(\alpha),
\end{equation}
where $\alpha=\sigma^2$ and 
$$
A=
\begin{pmatrix}
0 & 0 & 2\\
0 & 0 & 2\kappa^2\\
\kappa^2 & 1 & 0
\end{pmatrix}\ , \qquad
B=
\begin{pmatrix}
0 & 0 & 2\\
0 & 0 & -2\omega^2\\
-\omega^2 & 1 & 0
\end{pmatrix}\ , \qquad
J=
\begin{pmatrix}
0 & 0 & 0\\
1 & 0 & 0\\
0 & 0 & 0
\end{pmatrix}\ .
$$

If $\lambda_1$, $\lambda_2$, and $\lambda_3$ are the eigenvalues of $f(\alpha)$, then the distortion of the flow after $n$ periods is given by

\begin{equation}
    \beta(n)=\left[\frac{\text{min}\{|\lambda_1|,|\lambda_2|,|\lambda_3|\}}    {\text{max}\{|\lambda_1|,|\lambda_2|,|\lambda_3|\}}\right]^{n/2}\,.
\end{equation}

It is obvious that for $f(0)$ we have $\lambda_1=1$, $\lambda_2=\mu$, and $\lambda_3=\mu^{-1}$, where
\begin{itemize}
    \item $\mu$ is real and $0<\mu\le1$, or
    \item $\mu$ is complex and $|\mu|=1$,
\end{itemize}
and is given by
\begin{equation}
    \mu=\tfrac{1}{4}\left(b -\sqrt{b^2-4}\right)^2,
\end{equation}
where

\begin{equation}
    b=2 \cosh (\kappa\tau_\kappa) \cos (\omega \tau_\omega) 
    +\left(\frac{\kappa}{\omega } -\frac{\omega}{\kappa }\right)  \sinh (\kappa \tau_\kappa) 
    \sin (\omega \tau_\omega)\,,
\end{equation}
and

\begin{equation}
\text{tr}(f(0))=b^2-1\,.
\end{equation}

In addition, it can be shown that if $\sigma^2\left|\text{tr}\left(f'(0)\right)\right|\ll|b^2-4|$, then
\begin{equation}
    \beta(n)\approx\begin{cases}
    1-\frac{3}{2n}\sigma^2 
    \left|\frac{\text{tr}\left(f'(0)\right)}{{b^2-4}}\right|\,, & \text{if}\qquad |b|<2\\
    \mu^n &  \text{if}\qquad |b|>2
    \end{cases}\,.
\end{equation}

Hence, non-spherical distribution of photons ($\sigma^2>0$) contributes to the distortion as well.

For given $z$, $n$ is given by
\begin{equation}
n(z)=\frac{\sqrt{2}}{\tau_\kappa+\tau_\omega}\int_0^z\left[\Omega_m^{-1}-1+(1+\xi)^3\right]^{-1/2}d\xi\ .
\end{equation}

\section{Hyperbolicity signatures}

The hyperbolicity of the photon beams caused by observed parameters of underdense regions, voids, were shown to be compatible with the elongation of the excursion sets in temperature maps of CMB sky maps obtained by WMAP satellite \cite{G,GK2}. The signature of the deviation of the photon beams in voids was shown to fit the Kolmogorov stochasticity parameter map obtained for CMB temperature data in the Cold Spot region \cite{spot}. 

Another effect in which the described hyperbolicity can contribute is the distortion of the redshift-space in the galactic surveys defined by the correlation function of the separations of galaxies in line-of-sight and tangential directions \cite{Pea,Guz}.
That effect is attributed to the peculiar velocities of the galaxies within the galactic groups, clusters and superclusters, including the infall of galaxies to the cluster center (Kaiser effect), as well as to the gravitational shift - blue or red - due to the potential well of the particular structure and its peculiar motion with respect to us \cite{Mc,RG}. 

As an illustration, let us consider the survey of 10,000 galaxies in 300 Mpc distance (i.e. at redshift $z=0.8$) for which the distortion $\beta \simeq 0.7$ has been reported \cite{Guz}.  That distortion if attributed mainly to the tangential component of galactic separation, would correspond to a cumulative effect of e.g. $N=6$ line-of-sight voids of mean diameter $D=50Mpc$ and mean density parameters of the walls (of 4 Mpc mean diameter) and voids , $\tilde\delta_{\text{Wall}}=10\,, \tilde\delta_{\text{Void}}=-0.8$, respectively.

Quantitatively, the tangential distortion due to hyperbolicity depends on the angular distribution of the photon beams as shown in Fig.~1. It is seen that, at even slightly anisotropic beams the distortion can occur both at negative and positive matter mean densities.


\begin{figure}[h]
\caption{Tangential distortion at isotropic ($\sigma=0$) and 
non-isotropic ($\sigma=0.1$) distribution of photon beams.}
\centering
\includegraphics[]{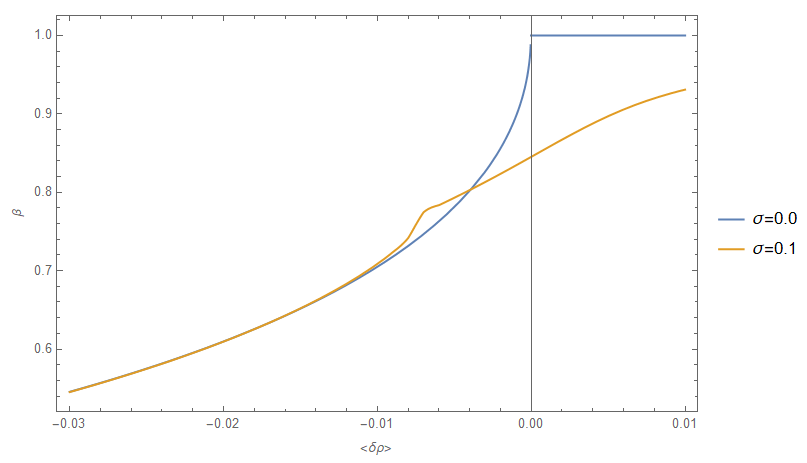}
\end{figure}

\section{Conclusion}

We considered the effect of hyperbolicity of photon beams at propagation through underdense regions, voids, known to be typical structures in the large scale matter distribution. Previously, the exponentially deviated beams have been associated with the anisotropy properties of the Cosmic Microwave Background temperature maps \cite{GK2}. The Cold Spot, the non-Gaussian region known in the CMB sky, had revealed properties peculiar to the hyperbolicity caused by a large void \cite{spot}; the conclusion on the void nature of the Cold Spot has been drawn also by 3D galactic survey \cite{Sz}.

Continuing the study of the signatures of the hyperbolicity of photon beams caused by voids, here we showed the dependence of the tangential distortion on the isotropy/anisotropy of the propagating beams. Namely, it appears that the distortion can occur both for negative or positive mean densities of matter, if the beam has a slight angular anisotropy. As an illustration, we mention the redshift-space distortion known for galactic surveys. Although the main contribution to that effect has to be due to the peculiar motion of galaxies in the groups and clusters, the considered effect of hyperbolicity can also have certain contribution; the latter issue needs thorough analysis with each given dataset.       


\end{document}